\documentclass[conference]{IEEEtran}

\usepackage{cite}
\usepackage{amsmath,amssymb,amsfonts}
\usepackage{graphicx}
\usepackage{booktabs}
\usepackage{stfloats}

\begin{document}

\title{WM4ISAC: World Model for Proactive ISAC Under Dynamic Blockage}

\author{\IEEEauthorblockN{Zichao Xiao$^\ast$, Linlong Wu$^\dag$, Bhavani Shankar M. R.$^\ast$}
\IEEEauthorblockA{{$^\ast$Interdisciplinary Centre for Security, Reliability and Trust (SnT), University of Luxembourg}}
\IEEEauthorblockA{{$^\dag$School of Information and Communication Engineering, University of Electronic Science and Technology of China}\\
Email: \{zichao.xiao, bhavani.shankar\}@uni.lu, linlong.wu@uestc.edu.cn}
}

\maketitle

\begin{abstract}
Dynamic blockage severely degrades vehicular integrated sensing and communication (ISAC), since narrow millimeter-wave beams must be adjusted before the line-of-sight (LoS) link is blocked or misaligned.
This paper proposes WM4ISAC, a proactive ISAC framework built on a world model that learns a reusable latent dynamics model of the traffic environment from downlink radar heatmaps, rather than directly mapping observations to beam indices.
The proposed framework uses a vision (V) module to compress sensing observations, an action-conditioned memory (M) module to predict latent environment evolution, and lightweight event heads to evaluate LoS alignment, blockage, and scatter-assisted non-line-of-sight (NLoS) alignment for candidate beams.
A codebook-level controller then uses these event predictions to proactively switch between LoS and NLoS beams without knowing the future user or blocker states.
Simulations under dynamic blockage show that WM4ISAC improves the mean rate by $28.9\%$ over current LoS tracking, raises the 5\%-tile rate from $0.9$ Mbps to $30.5$ Mbps, and reaches $88.7\%$ of the perfect-information upper bound.
\end{abstract}

\begin{IEEEkeywords}
World model, proactive ISAC, dynamic blockage, artificial intelligence (AI), machine learning (ML)
\end{IEEEkeywords}

\section{Introduction}

Integrated sensing and communication (ISAC) enables a base station (BS) to sense its surrounding environment while serving communication users, and is expected to support many emerging 6G applications \cite{zhang2021}.
A particularly important use case is sensing-assisted proactive beamforming for mobile millimeter-wave (mmWave) links, where narrow beams must be adjusted before severe misalignment occurs.
Compared with conventional pilot-based beam training, an ISAC BS can reuse the downlink communication waveform and its echoes to infer the environment evolution, thereby reducing pilot overhead and enabling proactive beam management \cite{liu2022integrated}.

The development of proactive ISAC is closely related to mmWave beam tracking and prediction, where external sensing sources such as radar, camera, lidar, GPS, or side information are often used to assist beam selection \cite{AA2018,Alrabeiah2020,Charan2021,vavutha2016,Graff2023,Demirhan2026}.
In pure ISAC systems, proactive beamforming can be roughly divided into model-based and learning-based methods.
For example, extended Kalman filtering (EKF) and message passing have been used to track and predict the user state for beam alignment \cite{liufan2020,yuan2021}.
Learning-based methods further explore data-driven mappings from historical channels or echoes to future beamformers, including unsupervised predictive beamforming \cite{liuchang2022}, transformer-based echo-to-beam mapping \cite{zhangyunwu2024}, motion-aware deep prediction \cite{zhangxiaoqi2024}, reinforcement-learning-based precoding \cite{yangzonghui2025}, and multi-modal feature fusion \cite{zhanghaotian2024}.

Despite these advances, dynamic blockage remains insufficiently addressed in proactive ISAC.
In vehicular scenarios, the line-of-sight (LoS) link of the served user may be blocked by other vehicles or roadside objects.
This not only causes a sharp communication-rate loss, but also makes sensing-based tracking more fragile and requires the BS to proactively switch to a non-line-of-sight (NLoS) backup link when necessary.
Some sensing-assisted beam tracking works have considered blockage prediction and multi-BS handoff \cite{AA2018,Alrabeiah2020,Charan2021}, but they often rely on additional sensing modalities or infrastructure and do not directly address how a single ISAC BS maintains the link using only its own communication echoes.
Moreover, many existing learning-based methods train a direct observation-to-action policy.
Such goal-oriented mappings can be effective for a fixed task, but they do not explicitly model how the physical environment evolves and how different beam actions affect future observations.
As a result, the learned representation is often tied to a specific decision objective, instead of exposing the underlying dynamics that could be reused for blockage prediction, LoS/NLoS switching, and future beam evaluation.

This motivates the use of world models for proactive ISAC.
The world model concept, popularized in \cite{ha2018world}, learns a compact latent dynamics model of the environment from spatio-temporal observations and actions.
By compressing high-dimensional observations into a latent space and learning how this latent state evolves, the model is encouraged to capture the compact physical regularities behind the observed motion and blockage process.
Instead of immediately outputting a task decision, it predicts how the environment will evolve under candidate actions, allowing the controller to reason about possible future consequences.
This idea has recently shown strong potential in game control, robotic policy evaluation, and predictive representation learning \cite{hafner2025training,quevedo2025worldgym,maes2026leworldmodel}.
For ISAC, the analogy is natural: sensing gives the BS ``eyes'' to observe the environment, while the recurrent hidden and cell states of a world model provide a compact ``memory'' of the learned dynamics for proactive decision-making.

In this paper, we propose WM4ISAC, standing for World Model for ISAC, as a proactive ISAC framework for dynamic blockage scenarios.
The considered roadside BS serves a moving communication user while using only the echoes of downlink communication signals, without extra sensing pilots after initialization.
The proposed world model follows a V-M structure: the V module encodes high-dimensional radar heatmaps into compact latent states, and the M module learns action-conditioned latent dynamics using a long short-term memory (LSTM) backbone with a mixture density network (MDN) output head.
The hidden and cell states of M summarize the learned environment dynamics, and lightweight event heads use this summary to predict LoS alignment, blockage, and NLoS alignment probabilities for candidate beams.
Finally, a codebook-level controller evaluates these candidates and proactively switches between LoS and scatter-assisted NLoS beams to maintain the communication link.

\section{System Model and Problem Formulation}
\label{sec:system_model}

\begin{figure}
    \centering
    \includegraphics[width=3in]{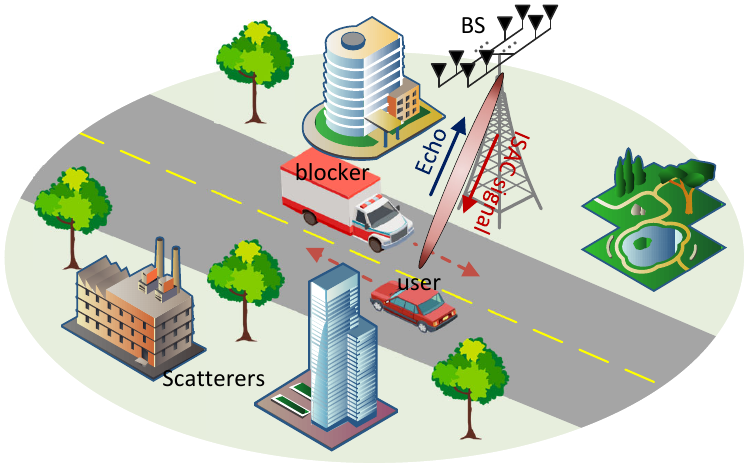} 
    \caption{The considered WM4ISAC application scenario.}
    \label{fig:considered_scenario} 
\end{figure}

As shown in Fig.~\ref{fig:considered_scenario}, we consider a monostatic roadside ISAC BS that serves a moving user while receiving the echoes of its downlink communication waveform.
The goal is to proactively update the communication beam before severe misalignment or blockage occurs.
A moving blocker may obstruct the LoS link, and a set of static scatterers can provide available NLoS backup paths.
After the initial uplink pilot, the BS only uses its own downlink echoes and the pre-stored static scatter map for online beam management.

The BS transmits orthogonal frequency-division multiplexing (OFDM) waveforms and is equipped with colocated transmit and receive uniform linear arrays (ULAs) with $N_\mathrm{t}$ and $N_\mathrm{r}$ antennas, respectively\footnote{We assume sufficient spacing with advanced full-duplex techniques \cite{besma2024}, and leakage self-interference is ignored in this work.}.
Time is divided into radar coherent processing intervals (CPIs), indexed by $k=0,1,\ldots,K-1$.
The user and blocker states are assumed to be approximately constant within each CPI.
At the $k$-th CPI, they are denoted by
\begin{equation}
\begin{aligned}
    \mathbf{s}_{\mathrm{u},k} &= [r_{\mathrm{u},k},\theta_{\mathrm{u},k},v_{\mathrm{u},k}]^T, \\
    \mathbf{s}_{\mathrm{b},k} &= [r_{\mathrm{b},k},\theta_{\mathrm{b},k},v_{\mathrm{b},k}]^T,
\end{aligned}
\end{equation}
where $r$, $\theta$, and $v$ denote range, angle of arrival (AoA), and radial velocity relative to the BS.
The static environment contains $Q$ scatterers.
The $q$-th scatterer is characterized by its AoA $\theta_q$, range $r_q$, and radar cross section (RCS) $\sigma_q$.
The static scatter map is defined as
\begin{equation}
    \mathcal{S}\triangleq \{(\theta_q,r_q,\sigma_q)\}_{q=1}^{Q}.
\end{equation}
We assume that $\mathcal{S}$ can be obtained and stored in advance through long-term background sensing or site calibration, and is available during online operation.

\subsection{Transmit Model}

The OFDM waveform has $N_\mathrm{s}$ subcarriers with spacing $\Delta f$ and carrier frequency $f_\mathrm{c}$.
Each CPI contains $L$ OFDM symbols.
Let $s_{k,l,i}$ denote the data symbol on the $i$-th subcarrier and the $l$-th OFDM symbol of CPI $k$, where $i=0,\ldots,N_\mathrm{s}-1$ and $l=0,\ldots,L-1$.
For notational simplicity, the symbol index $l$ is omitted when no ambiguity arises.
The transmitted baseband vector on subcarrier $i$ is
\begin{equation}
    \begin{aligned}
        \mathbf{x}_{k,i} = \mathbf{w}_{k} s_{k,i}, 
    \end{aligned}
\end{equation}
where $\mathbf{w}_{k}\in\mathcal{W}$ is the beamformer selected from a finite codebook $\mathcal{W}$ with $|\mathcal{W}|$ entries.

\subsection{Communication Model}
\label{subsec:comm_model}

After receive processing, the signal at the user is
\begin{equation}
    \begin{aligned}
        y_{k,i} = \mathbf{h}_{k,i}^H \mathbf{w}_{k} s_{k,i} + n_{k,i},
    \end{aligned}
\end{equation}
where $\mathbf{h}_{k,i}$ is the frequency-domain channel on subcarrier $i$ and is assumed constant within CPI $k$, and $n_{k,i}\sim \mathcal{CN}(0, \sigma_\mathrm{c}^2)$ is additive white Gaussian noise (AWGN).

Accordingly, the achievable communication rate of CPI $k$ is computed as
\begin{equation}
    \begin{aligned}
        \mathcal{R}_k = \sum_{i=0}^{N_\mathrm{s}-1} \Delta f
        \log_2 \left(1+\frac{|\mathbf{h}_{k,i}^H \mathbf{w}_{k}|^2}{\sigma_\mathrm{c}^2} \right).
    \end{aligned}
\end{equation}

\subsection{Sensing Model and Preprocessing}
\label{subsec:sens_model}

We model the user as a point target and the blocker as an extended target composed of multiple scattering points.
The total received echo is the superposition of the echoes from these points plus noise.
For a point with range $r$, AoA $\theta$, and radial velocity $v$, the corresponding echo component after receive processing at receive antenna $m=0,\ldots,N_\mathrm{r}-1$, subcarrier $i=0,\ldots,N_\mathrm{s}-1$, and OFDM symbol $l=0,\ldots,L-1$ is \cite{my2024}
\begin{equation}\label{eq:echo_model}
    \begin{aligned}
      \mathcal{Y}_k(m,i,l)
      &= \beta \mathbf{a}_\mathrm{t}(\theta)^H \mathbf{w}_{k} s_{k,i}
      e^{-\jmath 2\pi md_\mathrm{r}\sin \theta f_\mathrm{c}/c} \\
      &\quad \times
      e^{-\jmath 4\pi i \Delta f r/c}
      e^{\jmath 4\pi l T vf_\mathrm{c}/c}.
    \end{aligned}
\end{equation}
where $\beta$ accounts for path loss and RCS, $\mathbf{a}_\mathrm{t}(\theta)$ denotes the transmit steering vector, $d_\mathrm{r}$ is the receive-antenna spacing, $c$ is the speed of light, and $T$ is the OFDM symbol duration including CP.

Let $\widetilde{\mathcal{Y}}_k$ denote the composed echo cube at CPI $k$.
Since the transmitted data symbols are known at the BS, their modulation effect is removed by element-wise division, yielding $\widetilde{\mathcal{Y}}_k'$.
By applying a two-dimensional discrete Fourier transform (2D-DFT) to the spatial-fast-time and fast-slow-time slices, we obtain the angle-range and range-Doppler heatmaps, denoted by $\mathbf{H}_{\mathrm{ar},k}$ and $\mathbf{H}_{\mathrm{rd},k}$, respectively.
The BS observation is defined as
\begin{equation}
    \mathbf{o}_k \triangleq (\mathbf{H}_{\mathrm{ar},k},\mathbf{H}_{\mathrm{rd},k}).
\end{equation}
Since the static scatter map is known, the stationary background component is assumed to be filtered before forming $\mathbf{o}_k$.
Thus, the heatmaps mainly describe the dynamic user and blocker, while the static scatter information is provided separately through $\mathcal{S}$ for NLoS beam evaluation.

\subsection{Problem Formulation}

At the end of CPI $k$, the BS needs to select the beamformer for the next CPI using only historical sensing observations and the static scatter map.
Let
\begin{equation}
    \mathcal{O}_k \triangleq \{ \mathbf{o}_{k-O+1}, \ldots,\mathbf{o}_{k} \}
\end{equation}
be the observation history with window length $O$.
The proactive beam-selection problem is formulated as
\begin{equation} \label{eq:formulated_problem}
    \begin{aligned}
    \max_{\mathbf{w}_{k+1}}~&
    \mathbb{E}\!\left[
    \mathcal{R}_{k+1}(\mathbf{w}_{k+1})
    \mid \mathcal{O}_k,\mathcal{S}
    \right] \\
    \text{s.t.}~& \mathbf{w}_{k+1} \in \mathcal{W},
    \end{aligned}
\end{equation}
where the expectation is over the unknown future user/blocker states and channel.
This problem is challenging because the BS does not know the future communication channel, while the channel changes rapidly with user motion.
The moving blocker further makes the LoS link intermittently unavailable and can also confuse the sensing observations.
Even with the scatter map $\mathcal{S}$, the BS must infer when the LoS link will be blocked, which scatter-assisted NLoS path is useful, and when to switch back to LoS.

\section{Proposed WM4ISAC Framework}
\label{sec:wm4isac}

Instead of directly learning a black-box mapping from sensing observations to beam indices, WM4ISAC first learns a compact predictive model of the environment.
The learned dynamics are then converted into three action-conditioned events that are directly useful for beam selection: LoS alignment, blockage, and NLoS alignment.
Finally, a lightweight codebook-level controller evaluates all candidate beams through these event predictions.

\subsection{Latent World Model}

Following the world model idea in \cite{ha2018world}, WM4ISAC consists of a vision module (V) and a memory module (M).
The V module compresses the high-dimensional radar heatmap into a low-dimensional latent state, while the M module learns how this latent state evolves under beam actions.

We implement V as a variational autoencoder (VAE) \cite{vae}.
For each observation $\mathbf{o}_k$, the encoder produces the latent representation
\begin{equation}
    \mathbf{z}_k = E_{\phi}(\mathbf{o}_k),
    \label{vae_encoder}
\end{equation}
where $E_{\phi}(\cdot)$ is the VAE encoder with parameters $\phi$.

The M module is an action-conditioned MDN-LSTM.
Given the current latent state, the candidate next beam, and the recurrent memory, its LSTM backbone updates
\begin{equation}
    (\mathbf{h}_{k+1},\mathbf{c}_{k+1})
    =
    F_{\psi}(\mathbf{z}_k,\mathbf{w}_{k+1},\mathbf{h}_k,\mathbf{c}_k),
    \label{lstm_update}
\end{equation}
where $F_{\psi}(\cdot)$ denotes the LSTM transition with parameters $\psi$.
Here, $\psi$ denotes the parameters of the M module, including the LSTM backbone and the MDN head.
The hidden state $\mathbf{h}_k$ and cell state $\mathbf{c}_k$ summarize the past latent observations and actions.
Since complex physical environments can have multi-modal future evolution, M models the next latent state with a Gaussian mixture rather than a single deterministic prediction.
With $\mathbf{s}_k=(\mathbf{z}_k,\mathbf{h}_k,\mathbf{c}_k)$, this distribution is written as
\begin{align}
    p_{\psi}(\mathbf{z}_{k+1}|\mathbf{s}_k,\mathbf{w}_{k+1})
    &=
    \sum_{m=1}^{N_{\mathrm{mix}}}
    \pi_{m,k+1}
    \label{mdn_distribution}\\
    &\quad\times
    \mathcal{N}\!\left(
    \mathbf{z}_{k+1};
    \boldsymbol{\mu}_{m,k+1},
    \boldsymbol{\Sigma}_{m,k+1}
    \right),
    \notag
\end{align}
where $N_{\mathrm{mix}}$ is the number of mixture components, and
$\pi_{m,k+1}$, $\boldsymbol{\mu}_{m,k+1}$, and
$\boldsymbol{\Sigma}_{m,k+1}=\mathrm{diag}(\boldsymbol{\sigma}_{m,k+1}^2)$
are the mixture weight, mean, and diagonal covariance of the $m$-th component.
If M can predict how the latent environment changes after different beam actions, its recurrent states can be viewed as a compact memory of the learned physical dynamics.

\subsection{Event Heads and Codebook Controller}

The learned dynamics are used as an action evaluator for beam selection.
In a blockage-aware link, the controller needs to judge whether a candidate beam supports the LoS link, whether the LoS link will be blocked, and whether the candidate beam can exploit a useful scatter-assisted NLoS link.
We therefore attach three event heads to M.
For each candidate beamformer $\mathbf{w}\in\mathcal{W}$, the action-conditioned transition in \eqref{lstm_update} is first performed, and the event heads read the resulting state $(\mathbf{h}_{k+1},\mathbf{c}_{k+1})$ to output
\begin{equation}
    \begin{aligned}
    \hat p_l(\mathbf{w}) &{}= f_l(\mathbf{z}_k,\mathbf{w},\mathbf{h}_k,\mathbf{c}_k),\\
    \hat p_b(\mathbf{w}) &= P(b_{k+1}=1|\mathbf{z}_k,\mathbf{w},\mathbf{h}_k,\mathbf{c}_k),\\
    \hat p_g(\mathbf{w}) &{}= f_g(\mathbf{z}_k,\mathbf{w},\mathbf{h}_k,\mathbf{c}_k,\mathcal{S}),
    \end{aligned}
    \label{event_probabilities}
\end{equation}
where $b=1$ means that the LoS link is blocked, and $\hat p_l,\hat p_g\in[0,1]$ are LoS and NLoS beam-quality scores that predict the normalized qualities in \eqref{soft_labels}; only the NLoS head uses $\mathcal{S}$.
Since the heads read the post-action state, they evaluate the one-step consequence of each candidate beam, while the MDN head, trained by the likelihood term in \eqref{m_loss}, makes this state predictive of $\mathbf{z}_{k+1}$.

\begin{figure}
    \centering
    \includegraphics[width=\columnwidth]{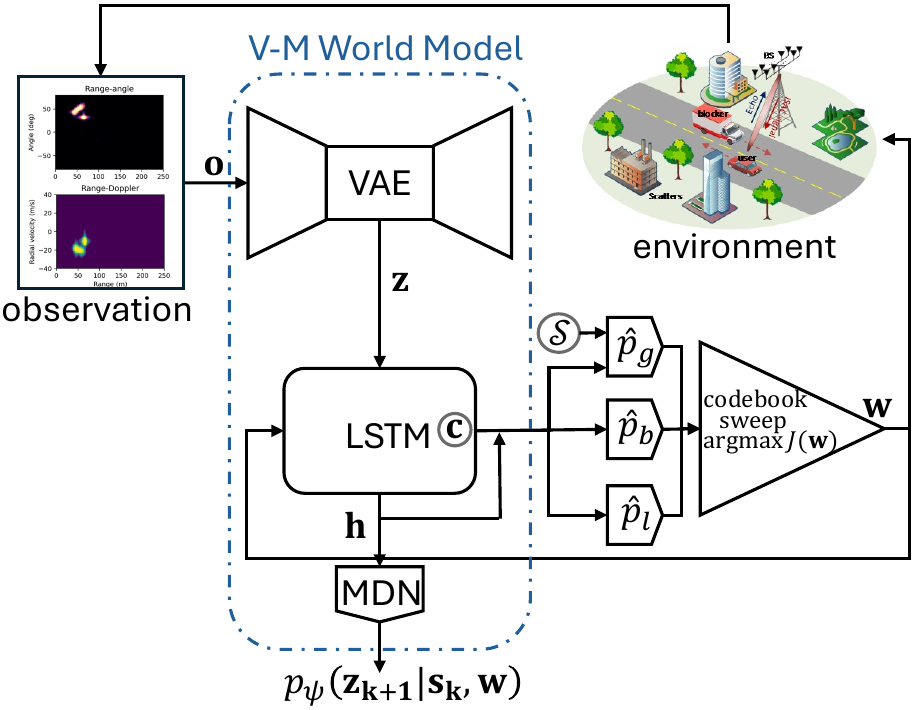} 
    \caption{The proposed WM4ISAC network structure.}
    \label{fig:wm4isac_structure} 
\end{figure}

The controller combines these outputs into a soft action score.
Since blockage is mainly a scene-level state, we aggregate its prediction over all candidate beams as
\begin{equation}
    \bar p_b = \frac{1}{|\mathcal{W}|}\sum_{\mathbf{w}\in\mathcal{W}}\hat p_b(\mathbf{w}).
    \label{aggregated_blockage}
\end{equation}
Then, the score of candidate beam $\mathbf{w}$ is defined as
\begin{equation}
    J(\mathbf{w}) = (1-\bar p_b)\hat p_l(\mathbf{w})+\bar p_b\hat p_g(\mathbf{w}).
    \label{controller_score}
\end{equation}
When blockage is unlikely, the first term dominates and the controller behaves as a LoS beam selector.
When blockage is likely, the second term dominates and the controller switches to the NLoS score.
$J(\mathbf{w})$ is a surrogate of \eqref{eq:formulated_problem}: $\mathbb{E}[\mathcal{R}_{k+1}]=(1-P_b)\mathbb{E}[\mathcal{R}_{k+1}|b{=}0]+P_b\mathbb{E}[\mathcal{R}_{k+1}|b{=}1]$, where the conditional rates increase with the LoS and NLoS beam qualities; \eqref{controller_score} replaces them by the predicted normalized qualities $\hat p_l$ and $\hat p_g$, and the blockage probability $P_b$ by $\bar p_b$.
Because the action space is a finite codebook, the next beam is selected by an exhaustive codebook sweep:
\begin{equation}
    \mathbf{w}_{k+1}^{\star}= \arg\max_{\mathbf{w}\in\mathcal{W}} J(\mathbf{w}).
    \label{controller_action}
\end{equation}
Thus, the controller is not a separately trained policy but directly uses the frozen world model as an action evaluator, as summarized in Fig.~\ref{fig:wm4isac_structure}.

\subsection{Training Objective}

The V and M modules are trained in two stages.
First, the V module compresses each radar heatmap observation $\mathbf{o}_k$ into a compact latent variable $\mathbf{z}_k$.
The encoder outputs a Gaussian posterior
$q_\phi(\mathbf{z}_k|\mathbf{o}_k)=\mathcal{N}(\boldsymbol{\mu}_{z,k},\mathrm{diag}(\boldsymbol{\sigma}_{z,k}^2))$, and the decoder reconstructs $\hat{\mathbf{o}}_k$.
The VAE is trained by
\begin{equation}
    \mathcal{L}_{\mathrm{V}}
    =
    \left\|\mathbf{o}_k-\hat{\mathbf{o}}_k\right\|_2^2
    +
    \lambda_{\mathrm{KL}}
    D_{\mathrm{KL}}
    \left(q_\phi(\mathbf{z}_k|\mathbf{o}_k)\|p(\mathbf{z}_k)\right),
    \label{vae_loss}
\end{equation}
where $p(\mathbf{z}_k)=\mathcal{N}(\mathbf{0},\mathbf{I})$ is the latent prior, and
$D_{\mathrm{KL}}(\cdot\|\cdot)$ denotes the Kullback--Leibler (KL) divergence that regularizes the posterior to stay close to this prior.
$\lambda_{\mathrm{KL}}$ balances reconstruction accuracy and latent regularization.
After V is trained, all heatmaps are encoded into latent sequences and V is kept fixed.

Second, M is trained on action-conditioned latent transitions.
For each transition, M receives $(\mathbf{z}_k,\mathbf{w}_{k+1})$, updates the LSTM state, and predicts the distribution of $\mathbf{z}_{k+1}$ through its MDN head.
This dynamics part is trained by the negative log-likelihood
$-\log p_\psi(\mathbf{z}_{k+1}|\mathbf{z}_k,\mathbf{w}_{k+1},\mathbf{h}_k,\mathbf{c}_k)$,
where $p_\psi(\cdot)$ is the MDN output distribution parameterized by M.

The event heads are supervised according to the semantics in \eqref{event_probabilities}.
The blockage target $y_b$ is the physical binary blockage state.
For the LoS and NLoS action-quality heads, we avoid hard top-1 labels because they discard the ordering among non-best beams.
Instead, their targets are normalized beam qualities of the true scene at CPI $k+1$:
\begin{equation}
    \begin{aligned}
    y_l(\mathbf{w}) &= \big(G_l(\mathbf{w})/\textstyle\max_{\mathbf{w}'\in\mathcal{W}}G_l(\mathbf{w}')\big)^{\gamma_l},\\
    y_g(\mathbf{w}) &= \big(G_g(\mathbf{w})/\textstyle\max_{\mathbf{w}'\in\mathcal{W}}G_g(\mathbf{w}')\big)^{\gamma_g},
    \end{aligned}
    \label{soft_labels}
\end{equation}
where $G_l(\mathbf{w})=|\mathbf{a}_\mathrm{t}(\theta_{\mathrm{u},k+1})^H\mathbf{w}|^2$ is the LoS gain, $G_g(\mathbf{w})=\max_{q\in\mathcal{Q}(\mathbf{w})}|\mathbf{a}_\mathrm{t}(\theta_q)^H\mathbf{w}|^2\sigma_q/(r_q^2d_{q,k+1}^2)$ is the single-bounce NLoS strength, $d_{q,k+1}$ is the scatterer-user distance, and $\mathcal{Q}(\mathbf{w})$ contains the two scatterers maximizing $|\mathbf{a}_\mathrm{t}(\theta_q)^H\mathbf{w}|^2\sigma_q/r_q^2$.
Thus, near-optimal beams receive higher labels than poor beams, making $\hat p_l$ and $\hat p_g$ useful for the online codebook sweep.

The resulting M objective is
\begin{align}
    \mathcal{L}_{\mathrm{M}}
    =
    &-\log p_\psi(\mathbf{z}_{k+1}|\mathbf{z}_k,\mathbf{w}_{k+1},\mathbf{h}_k,\mathbf{c}_k)
    \label{m_loss}\\
    &+
    \lambda_l\mathrm{BCE}(y_l,\hat p_l)
    +
    \lambda_b\mathrm{BCE}(y_b,\hat p_b)
    +
    \lambda_g\mathrm{BCE}(y_g,\hat p_g),
    \notag
\end{align}
where BCE denotes binary cross-entropy with hard or soft targets, and
$\lambda_l$, $\lambda_b$, and $\lambda_g$ weight the three event-supervision terms.
The first term learns the predictive world dynamics, while the event terms make the learned state directly usable for evaluating candidate beams in \eqref{controller_score}.

\section{Simulations}

\subsection{Simulation Setup}

We evaluate WM4ISAC in the dynamic-blockage scenario described in Section~\ref{sec:system_model}.
The simulation section has three purposes: to quantify the online rate gain, to show how the proposed controller behaves during a blockage event, and to test whether the learned dynamics remain useful when the blockage severity changes.
Unless otherwise specified, the main parameters are listed in Table~\ref{tab:sim_params}.
The communication channel follows the geometric wideband mmWave model, with one LoS component and single-bounce components through static scatterers.
The same physical scene is used to generate both the sensing echoes and the subcarrier channels for rate evaluation.
Consistent with the observation model, the static background is filtered from the radar heatmaps, while the scatter map $\mathcal{S}$ is separately available for NLoS beam evaluation.

For world model training, we generate $10240$ training scenes and $2560$ validation scenes.
Each physical scene is simulated with $11$ action rollouts: $7$ random exploratory rollouts and $4$ guided/oracle-like rollouts covering LoS following, motion extrapolation, LoS/NLoS switching, and NLoS-only behavior.
These rollouts share the same target route, blocker trajectory, and scatter map, but use different beam actions.
Therefore, M learns action-conditioned latent dynamics rather than imitating a single behavior policy.
The best validation checkpoint is used for all online tests.

\begin{table}[t]
    \centering
    \caption{Main simulation and training parameters.}
    \label{tab:sim_params}
    \small
    \setlength{\tabcolsep}{5pt}
    \begin{tabular}{lc}
        \toprule
        Parameter & Value \\
        \midrule
        Carrier frequency & $28$ GHz \\
        BS transmit/receive antennas & $64/64$ \\
        OFDM subcarriers / spacing & $256$ / $120$ kHz \\
        OFDM symbols per CPI & $64$ \\
        Transmit/noise power & $1$ W / $10^{-10}$ W \\
        CPI interval & $0.1$ s \\
        CPI steps per episode & $120$ \\
        Observation window length $O$ & $120$ CPIs \\
        Steering codebook entries & $64$ \\
        Target speed & $16$--$30$ m/s \\
        Blocker speed / width & $8$--$24$ m/s / $2.5$--$6$ m \\
        Effective blockage-zone length & $14$--$48$ m \\
        Complete blockage duration & $5$--$30$ CPIs \\
        Static scatterers & $24$ \\
        VAE latent dimension & $64$ \\
        M hidden dimension / MDN mixtures & $1024$ / $5$ \\
        Local NLoS scatter tokens & $2$ \\
        Event-label exponents $\gamma_l/\gamma_g$ & $2/2$ \\
        VAE training & $40$ epochs, batch $12288$ \\
        M training & $200$ epochs, batch $1024$ \\
        \bottomrule
    \end{tabular}
\end{table}

\subsection{Online Rate Evaluation}

For online evaluation, we compare ``Proposed WM4ISAC'' with three baselines.
The first baseline, ``Current LoS beam'', selects the beam nearest to the currently observed target direction without predicting future motion or switching to NLoS paths.
The second baseline, ``Perfect NLoS beam'', always selects the scatter-assisted backup beam maximizing $G_g(\mathbf{w})$ in \eqref{soft_labels} using perfect scene information.
It is used as a blockage-period reference, but it is not expected to be optimal when LoS is available.
The third baseline, ``Perfect-info upper bound'', switches between the target-facing LoS beam and the best NLoS beam using the true scene state, and serves as an ideal upper bound.
Learning-based predictive beamforming \cite{liuchang2022,zhangyunwu2024,zhangxiaoqi2024} predicts the next LoS beam and is not directly applicable here, since the user echo vanishes during blockage and no NLoS backup beam is selected; a comparison with learned observation-to-beam baselines trained for blockage is left for future work.
All methods are evaluated online on $2048$ independent scenes with randomized target routes, blocker trajectories, blockage intervals, and scatter maps.
The reported rate is always the true simulator rate, not the predicted world model score.
In Table~\ref{tab:main_online_results}, ``Blocked'' reports the mean rate over blocked CPIs, ``5\%-tile'' measures the low-rate tail over all CPIs, and ``Mean/UB'' denotes the ratio between each method's mean rate and the mean rate of the perfect-information upper bound.

\begin{table}[t]
    \centering
    \caption{Online performance on $2048$ held-out scenes.}
    \label{tab:main_online_results}
    \begin{tabular}{lcccc}
        \toprule
        Method & Mean & Blocked & 5\%-tile & Mean/UB \\
        & Mbps & Mbps & Mbps & \\
        \midrule
        Current LoS tracking & 110.5 & 13.0 & 0.9 & 0.688 \\
        Perfect NLoS beam & 90.8 & 88.1 & 16.1 & 0.566 \\
        \textbf{Proposed WM4ISAC} & \textbf{142.4} & 76.7 & \textbf{30.5} & \textbf{0.887} \\
        Perfect-info upper bound & 160.5 & 88.1 & 67.4 & 1.000 \\
        \bottomrule
    \end{tabular}
\end{table}

Table~\ref{tab:main_online_results} first quantifies the overall online rate performance.
The current LoS baseline achieves $110.5$ Mbps mean rate, but its blocked-CPI mean rate drops to only $13.0$ Mbps.
This confirms that current-direction tracking alone is not sufficient in a narrow-beam mobile link: the beam must be proactive and blockage-aware.
WM4ISAC increases the mean rate to $142.4$ Mbps, corresponding to a $28.9\%$ gain over current LoS tracking, and improves the 5\%-tile rate from $0.9$ Mbps to $30.5$ Mbps.
During blocked CPIs, WM4ISAC achieves $76.7$ Mbps, close to the $88.1$ Mbps blockage reference of the perfect NLoS beam.
However, the perfect NLoS beam has only $90.8$ Mbps overall mean rate because it wastes the strong LoS link when the user is not blocked.
Therefore, the benefit of WM4ISAC is the learned switching behavior: it uses LoS when available and switches to scatter-assisted NLoS when blockage is likely.
Overall, WM4ISAC reaches $88.7\%$ of the perfect-information upper bound without access to the true future target or blockage states.

\begin{figure}[t]
    \centering
    \includegraphics[width=0.8\columnwidth]{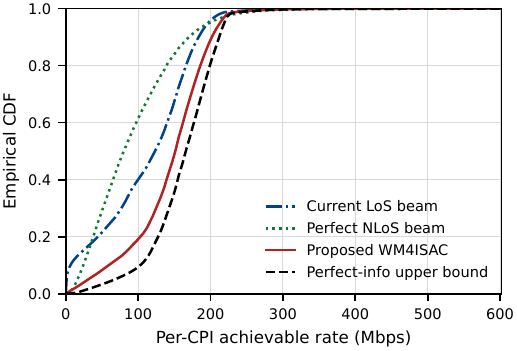}
    \caption{Empirical CDF of the per-CPI achievable communication rate.}
    \label{fig:rate_cdf}
\end{figure}

To further examine reliability, Fig.~\ref{fig:rate_cdf} plots the empirical cumulative distribution function (CDF) of the per-CPI achievable rate over all evaluated CPIs.
The current LoS curve has a pronounced low-rate tail, which is consistent with its small 5\%-tile rate in Table~\ref{tab:main_online_results}.
The perfect NLoS curve reduces part of this blockage-induced tail, but its rate distribution is shifted downward in many LoS-clear CPIs.
In contrast, WM4ISAC pushes the low-rate tail to the right while remaining close to the perfect-information upper bound.
This shows that the proposed controller improves both average throughput and service continuity.

\begin{figure}[t]
    \centering
    \includegraphics[width=\columnwidth]{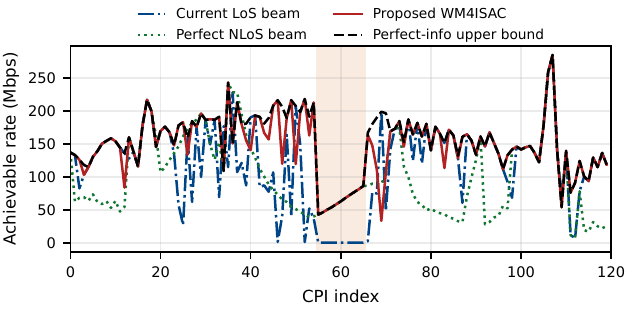}
    \caption{Representative online episode with a dynamic blockage interval.}
    \label{fig:representative_episode}
\end{figure}

Fig.~\ref{fig:representative_episode} illustrates the same mechanism in one representative episode.
Outside the blockage interval, WM4ISAC mostly tracks the LoS link proactively and stays close to the perfect-information upper bound, with a few short drops caused by one-beam misalignment.
When blockage occurs, current LoS tracking collapses because it continues to point toward the blocked direction, whereas WM4ISAC switches to the scatter-assisted NLoS path.
After the blockage ends, WM4ISAC returns to the LoS beam instead of remaining on the backup link.

\subsection{Blockage-Severity Evaluation}

\begin{figure}[t]
    \centering
    \includegraphics[width=0.8\columnwidth]{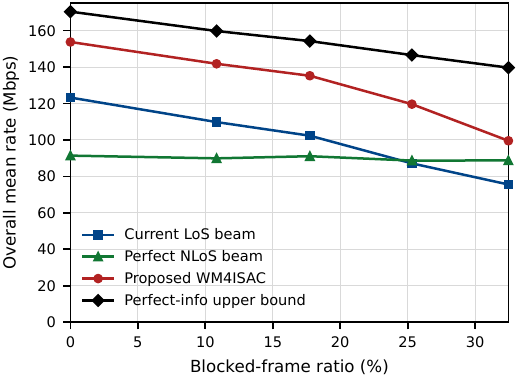}
    \caption{Mean rate versus the realized blocked-frame ratio.}
    \label{fig:blockage_severity}
\end{figure}

Finally, Fig.~\ref{fig:blockage_severity} tests robustness to different blockage severities.
For this experiment, we evaluate $512$ scenes per setting and change only the blocker size and motion range; the other parameters follow Table~\ref{tab:sim_params}.
The horizontal axis is the realized blocked-frame ratio, i.e., the fraction of CPIs in which the LoS link is blocked.
When the ratio is zero, WM4ISAC reaches $153.8$ Mbps, higher than the $123.3$ Mbps of current LoS tracking, showing that proactive beam adjustment is useful even without blockage.
As the realized blocked-frame ratio increases from $10.8\%$ to $32.5\%$, the mean rate of current LoS tracking decreases from $109.8$ Mbps to $75.5$ Mbps, while WM4ISAC decreases from $141.8$ Mbps to $99.5$ Mbps.
Under the most severe setting, WM4ISAC still exceeds the perfect NLoS beam baseline ($88.8$ Mbps), indicating that it exploits the NLoS path during blockage while recovering the LoS link afterward.
This graceful degradation shows that the learned dynamics generalize beyond the nominal blockage setting used for training.

\section{Conclusion}
\label{sec:conclusion}

In this paper, we proposed WM4ISAC, a proactive ISAC framework that learns compact latent environment dynamics from radar heatmaps for beam management under dynamic blockage.
By separating dynamics learning from beam selection, WM4ISAC uses V to compress sensing observations, M to predict action-conditioned latent evolution, and event heads to evaluate LoS, blockage, and NLoS beam usefulness.
Simulation results show that WM4ISAC improves the mean rate by $28.9\%$ over current LoS tracking, raises the low-rate tail, and reaches $88.7\%$ of the perfect-information upper bound, demonstrating the value of learned world dynamics for proactive blockage-aware ISAC.

\bibliographystyle{IEEEtran}
\bibliography{refs}

\end{document}